\documentclass[aps,prb,amsmath,10pt,twocolumn,amssymb,titlepage,superscriptaddress]{revtex4-1}

\usepackage[english]{babel}
\usepackage{calc}

\usepackage{graphicx}

\usepackage{nicefrac}
\usepackage{amsfonts}

\usepackage{amssymb}
\usepackage{amsmath} 

\usepackage{subfigure}
\usepackage{multirow} 
\usepackage{tabularx} 
\usepackage{array}

\usepackage{units}

\usepackage{tensor} 
\usepackage{braket}

\usepackage{bm}
\usepackage{hyperref}

\usepackage{soul}  

\usepackage{color} 
\usepackage{romannum}
\usepackage{float}

\usepackage{sidecap}
\sidecaptionvpos{figure}{c}
\graphicspath{{figures/}}

\begin{document}

\title{Machine-learned multi-system surrogate models for materials prediction}

\author{Chandramouli Nyshadham}
\affiliation{Department of Physics and Astronomy, Brigham Young University, Provo, UT 84602, USA}

\author{Matthias Rupp}
\affiliation{Fritz Haber Institute of the Max Planck Society, Faradayweg 4--6, 14195 Berlin, Germany}
\affiliation{Present address: Citrine Informatics, 702 Marshall Street, Redwood City, CA 94063, USA}

\author{Brayden Bekker}
\affiliation{Department of Physics and Astronomy, Brigham Young University, Provo, UT 84602, USA}

\author{Alexander V. Shapeev}
\affiliation{Skolkovo Institute of Science and Technology, Skolkovo Innovation Center, Building 3, Moscow, 143026, Russia}

\author{Tim Mueller}
\affiliation{\mbox{Department of Materials Science and Engineering, Johns Hopkins University, Baltimore, MD~21218, USA}}

\author{Conrad W. Rosenbrock}
\affiliation{Department of Physics and Astronomy, Brigham Young University, Provo, UT 84602, USA}

\author{G\'abor Cs\'anyi}
\affiliation{\mbox{Engineering Laboratory, University of Cambridge, Trumpington Street, Cambridge CB2~1PZ, United Kingdom}}

\author{David W. Wingate}
\affiliation{Computer Science Department, Brigham Young University, Provo, UT 84602, USA}

\author{Gus L. W. Hart}
\affiliation{Department of Physics and Astronomy, Brigham Young University, Provo, UT 84602, USA}

\begin{abstract}\noindent
Surrogate machine-learning models are transforming computational materials science by predicting properties of materials with the accuracy of ab initio methods at a fraction of the computational cost.
We demonstrate surrogate models that simultaneously interpolate energies of different materials on a dataset of 10 binary alloys (AgCu, AlFe, AlMg, AlNi, AlTi, CoNi, CuFe,~CuNi,~FeV,~NbNi) with 10 different species and all possible fcc, bcc and hcp structures up to 8 atoms in the unit cell, 15\,950 structures in total.
We find that the deviation of prediction errors when increasing the number of simultaneously modeled alloys is less than 1\,meV/atom. 
Several state-of-the-art materials representations and learning algorithms were found to qualitatively agree on the prediction errors of formation enthalpy with relative errors of  $<$2.5\% for all systems.
\end{abstract}

\maketitle


\section*{introduction}

Advances in computational power and electronic structure methods have enabled large materials databases \cite{cswxytnhsbml2012,skamw2013,johcrdcgscp2013,nomadinterview}.
Using high-throughput approaches,\cite{chbnmsl2013} these databases have proven a useful tool to predict the properties of materials.
However, given the combinatorial nature of materials space,\cite{iotgct2017,w2015b} it is infeasible to compute properties for more than a tiny fraction of all possible materials using electronic structure methods such as density functional theory\cite{hk1964,ks1965} (DFT).
A potential answer to this challenge lies in a new paradigm: surrogate machine-learning models for accurate materials predictions\cite{bpkc2010,s2016,hr2017}.

\begin{figure*}[bhtp]
	\begin{minipage}[b]{0.575\linewidth}
		\raisebox{1ex-\height}{\includegraphics[width=\linewidth]{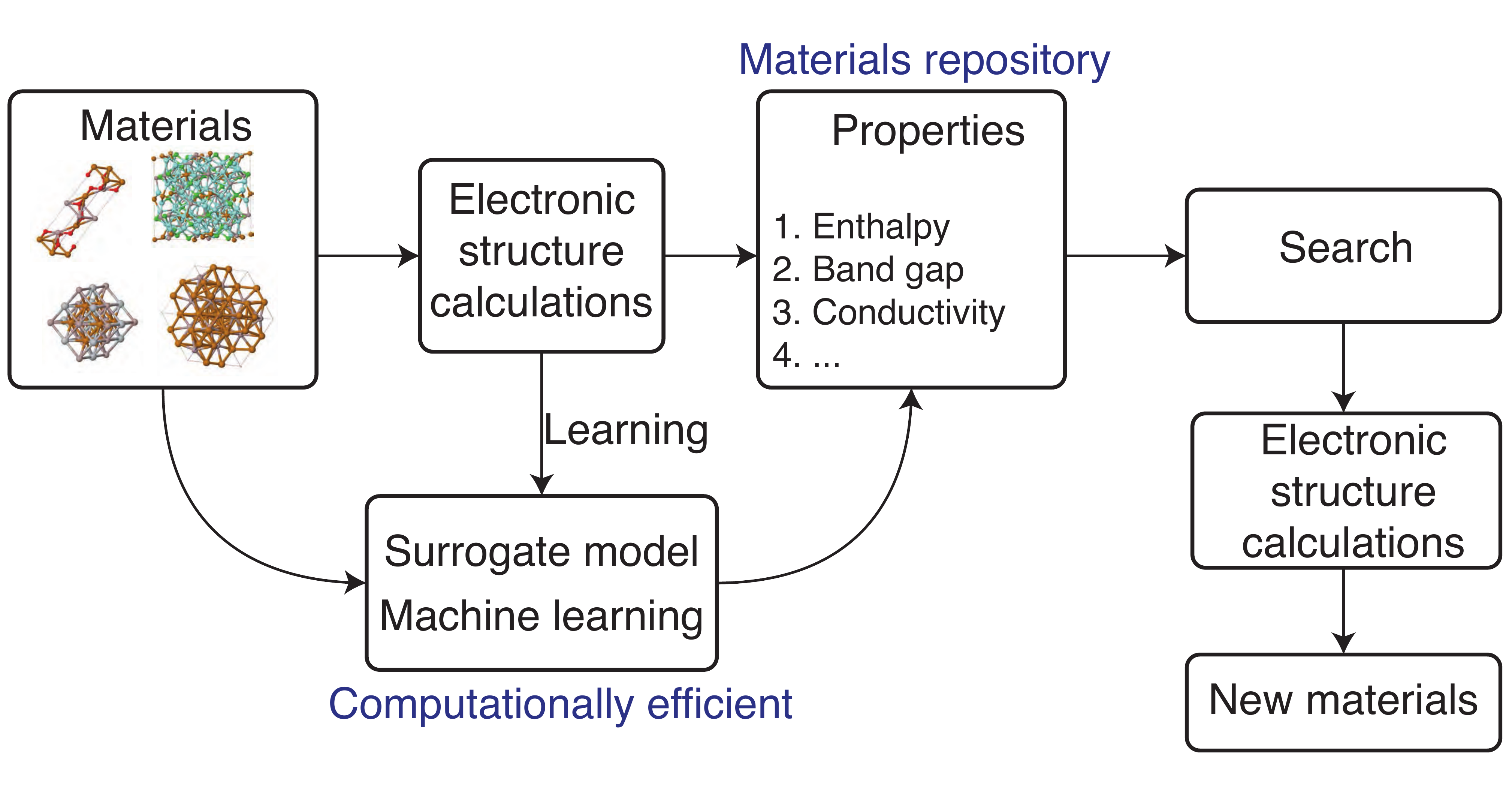}}\\ \mbox{}
	\end{minipage}\hfill%
	\begin{minipage}[b]{0.425\linewidth-2.5ex}
		\caption{\textit{The accelerated high-throughput approach.}
			Candidate structures and properties are generated by surrogate machine-learning models based on reference electronic structure calculations in a materials repository.
			Selected structures are validated by electronic structure calculations, preventing false positive errors.
			\label{AHT}}
			
			\mbox{}
	\end{minipage}
\end{figure*}

The key idea is to use machine learning to rapidly and accurately interpolate between reference simulations,
effectively mapping the problem of numerically solving for the electronic structure of a material onto a statistical regression problem\cite{r2015}. 
Such fast surrogate models could be used to filter the most suitable materials from a large pool of possible materials and then validate the found subset by electronic structure calculations. 
Such an ``accelerated high-throughput" (AHT) approach (Figure~\ref{AHT}) could potentially increase the number of investigated materials by several orders of magnitude.

Traditionally, empirical interatomic potentials were used to  reproduce  macroscopic properties of materials faster than DFT. Well-known  empirical interatomic potentials for periodic solids include Lennard-Jones potentials, the Stillinger-Weber potential and Embedded Atom Methods (EAM) for alloys. A problem with empirical interatomic potentials is that they are designed with a fixed functional form and cannot be systematically improved. In contrast, surrogate models which are empirical interatomic models based on machine learning systematically improve with additional data. This  potential advantage over traditional potentials has resulted in the proposal of many machine-learned surrogate models for materials prediction.


We demonstrate the feasibility of  machine-learned surrogate models  for predicting  enthalpies of formation of materials across composition, lattice types, and atomic configurations. Our findings were motivated  towards knowing whether different surrogate models proposed in the literature are consistent in their predictions of formation enthalpy rather than comparing the performance of different surrogate models.
We find that five combinations of state-of-the-art representations and regression methods (Table~\ref{tab:SurMod}) all yield consistent predictions with errors of $\sim$10 meV/atom or less depending on the system. We also find that when we combined the data from all 10 systems to build a single model; the combined model is essentially as good as the 10 individual models.

\begin{table*}[t]
	\caption{\textit{State-of-the-art surrogate machine-learning models} investigated in this work.
	\label{tab:SurMod}}
	
	\smallskip
	
	\begin{tabular}{@{}p{0.07\linewidth}p{0.4\linewidth}p{0.53\linewidth-4\tabcolsep}@{}}
		\hline
		Abbrv. & Surrogate model \rule{0pt}{11pt} & Description \\[0.5ex]
		\hline
		
		CE \rule{0pt}{11pt} &
		Cluster Expansion \cite{sdg1984,f1994,wc2002,mc2009b} + Bayesian approach\cite{mc2009b} & 
		One of the early successful surrogate models developed in the materials community. 
		A material's ground state energy is expanded as an Ising-type model with constant expansion coefficients.  
		\\[1ex]
		
		MBTR\linebreak +KRR & \raggedright%
		Many-Body Tensor Representation \cite{hr2017} + \linebreak Kernel Ridge Regression &  
		Materials are expanded in distributions of $k$-body terms stratified by chemical element species,
		using non-linear regression.
		\\[1ex]
		
		MBTR\linebreak +DNN & \raggedright%
		Many-Body Tensor Representation + \linebreak Deep Neural Network &  
		MBTR is  used as input for DNN to  learn a new representation and predict  using a parametric deep regression method.\linebreak
		\\[1ex]
		
		SOAP\linebreak +GPR & \raggedright%
		Smooth Overlap of Atomic Positions \cite{bkc2013} + \linebreak Gaussian Process Regression \cite{bpkc2010} &  
		Atomic environments represented as smoothened Gaussian densities of neighboring atoms expanded in a spherical harmonics basis,
		using non-parametric regression.
		\\[1ex]
		
		MTP & \raggedright%
		Moment Tensor Potentials (MTP) \cite{s2016} + \linebreak Polynomial Regression &  
		Atomic environments expanded in a tailored polynomial basis, computed via contractions of moment tensors.
		\\[1ex]
	
	\hline
	\end{tabular}
\end{table*}



A surrogate machine-learning model replaces ab initio simulations by mapping a crystal structure to properties such as formation enthalpy, elastic constants, or band gaps, etc.
Their  utility lies in the fact that once the model is trained, properties of new materials can be predicted very quickly. The prediction time is either constant, or scales linearly with the number of atoms in the system,  with a low pre-factor, typically in milliseconds.

The two major parts of a surrogate machine-learning model are the numerical representation of the input data \cite{sgbsmg2014,s2016} and the learning algorithm.
We use the term ``representation" for a set of features (as opposed to a collection of unrelated or only loosely related descriptors) that satisfies certain physical requirements \cite{hr2017,r2015,bkc2013,lrrk2015} such as invariance to translation, rotation, permutation of  atoms, uniqueness (representation is variant against transformations changing the property, as systems with identical representation but differing in the property would introduce errors\cite{m2012e}), differentiability, and computational efficiency.
The role of the representation is akin to that of a basis set in that the predicted property is expanded in terms of a set of reference structures. 

To model materials, it is desirable that a representation enables accurate predictions and is able to handle multiple elements simultaneously. The materials community has proposed several representations \cite{bp2007,b2011b,sgbsmg2014,b2014,bkc2013,s2016,bpkc2010,flla2015,hr2017} for crystal structures. 
Some do not fulfill the above properties exactly or are restricted \emph{in practice} to materials with a single element. Consequently, surrogate models based on these representations are limited in their accuracy, due to the violation of any of the physical requirements mentioned above (e.g., for the sorted and eigenspectrum variants of the Coulomb matrix, continuity and uniqueness, respectively\cite{m2012e,lrrk2015}).

We explore three state-of-the-art representations that fulfill above properties for construction of general surrogate models:
Many-body tensor representation \cite{hr2017} (MBTR), smooth overlap of atomic positions \cite{bpkc2010,bkc2013} (SOAP) and moment tensor potentials \cite{s2016} (MTP).
Each representation is employed as proposed and implemented by its authors, including the regression method:
Kernel ridge regression\cite{r2015} (KRR) for MBTR, Gaussian process regression regression \cite{rw2006} (GPR) for SOAP, and polynomial regression \cite{s2016} for MTP.
Since predictions (but not necessarily other properties) of the kernel-based KRR and GPR are identical, we will use the two terms interchangeably here.
We also employed cluster expansion \cite{sdg1984,f1994,wc2002,mc2009b} (CE) and Deep Neural Network \cite{lbh2015,s2015b} (DNN) models. Our purpose is not to compare the performance of these different surrogate models. Consequently, the models were not optimized to minimize the error; rather they were generated to maintain a typical speed/accuracy balance.

CE models have been used for three decades to efficiently model ground state energies of metal alloys, but require that the atomic structure can be mapped to site occupancies on a fixed lattice.
They are therefore less suited to model different materials. 
In this work, we use them as a baseline and build a separate CE model for each alloy. The comparison is not between CE and other models regarding performance, but our intention is to see how consistent are these different models in predicting the formation enthalpy of materials.

DNNs are essentially recursively stacked layers of functions, a large number of layers being a major difference between DNNs and conventional neural networks. 
They have been used to predict energies \cite{ssktm2018,lsb2018,mst2017,sir2017,sacmt2017} and to learn representations\cite{mds2018,gd2017}. 
While DNNs can learn representations (``end-to-end learning'', here from nuclear charges, atom positions and unit cell basis vectors to enthalpy of formation), this requires substantially more data than starting with a representation as input\cite{bp2007,b2011b,b2014}. 
We, therefore, provide the DNN with MBTR as input. MBTR is a manually designed representation and works well with the Gaussian kernel.
The idea of using MBTR along with DNN is to explore whether a representation-learning technique can improve upon a manually designed representation in conjunction with the standard Gaussian kernel (MBTR+KRR).


\section*{Results and discussion}

\subsection*{Energy predictions for single alloys}

\begin{figure*}
    \centering
    \includegraphics[scale=0.6]{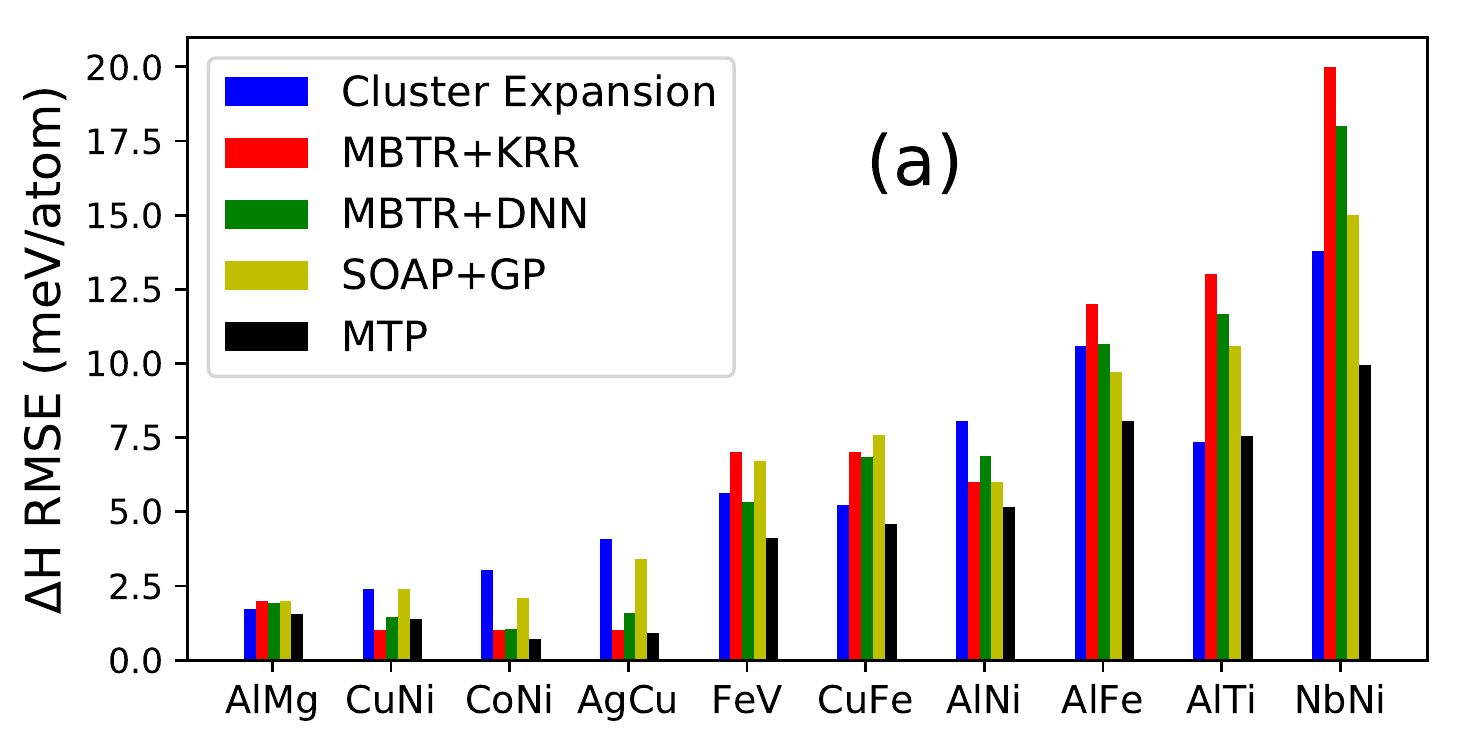}
    \includegraphics[scale=0.5]{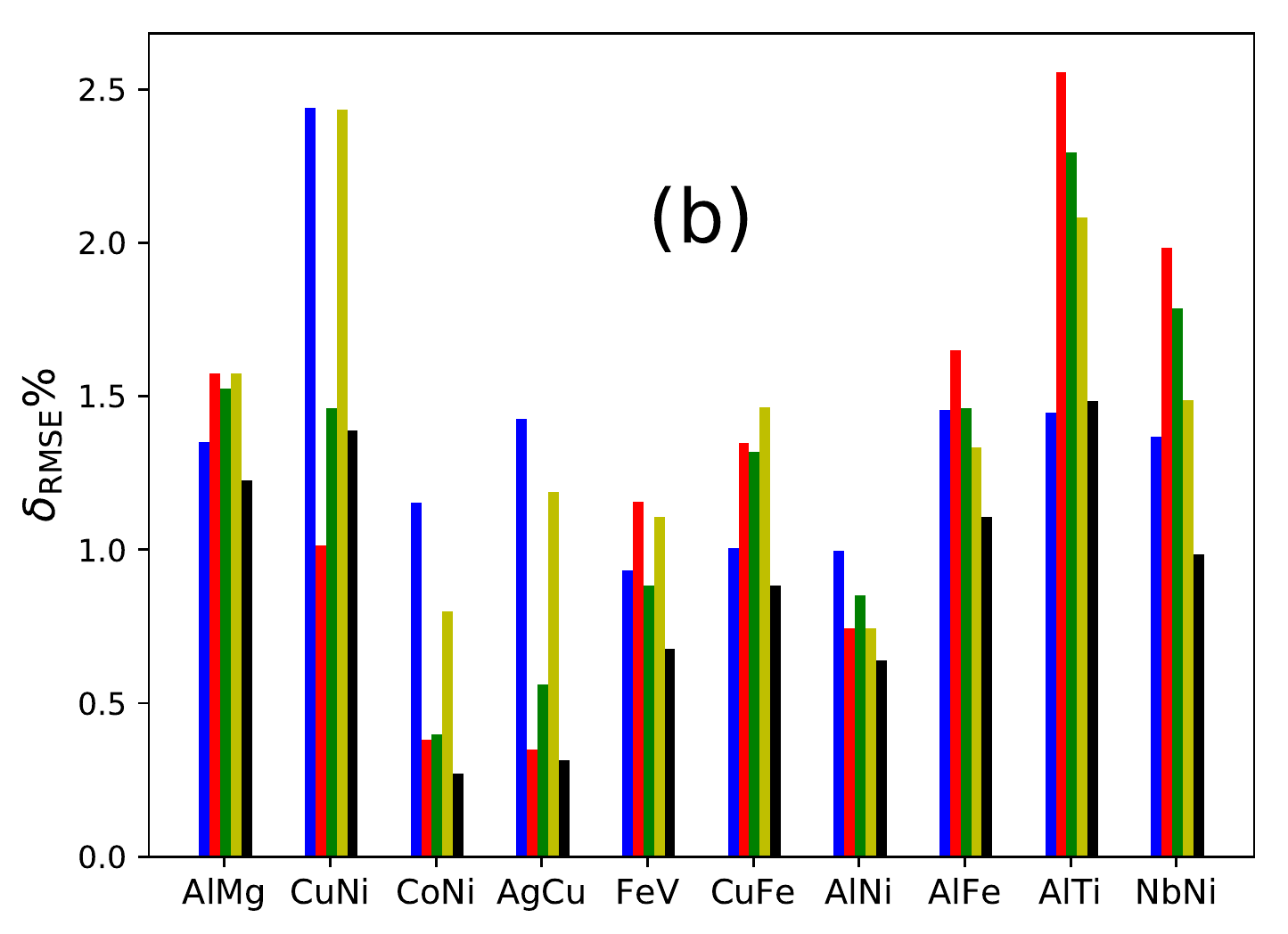}
    \caption{\textit{Consistency in prediction errors of formation enthalpy} of five machine learning surrogate models on the \texttt{DFT-10B} dataset. (a) Root mean squared error (RMSE) of predicted enthalpies of formation of each surrogate model on each binary alloy subset in meV/atom (colored bars).  RMSE for MTP results is computed using pure atom total energies obtained from DFT. The consistency of errors across models indicates the validity of machine learning surrogate models to predict formation enthalpy of materials---prediction errors are similar, independent of the details of model parametrization. (b) Root mean squared error (RMSE) of predicted enthalpies of formation of each surrogate model on each binary alloy subset as a percentage of energy range.
		Note that relative errors are below 2.5\% for all systems.
     }\label{fig:allModels}
\end{figure*}

Prediction errors for enthalpies of formation of each of the five surrogate models on each binary alloy subset of the data are presented in Figure~\ref{fig:allModels}a.
Prediction errors of all surrogate models agree qualitatively on all subsets of the data.
We interpret this consistency  to be indicative of the validity of the machine learning approach to surrogate models of formation enthalpy of materials, independently of the parametrization details of the models.

For four binary systems (AgCu, AlMg, CoNi, CuNi) predictions errors are below 3\,meV/atom. The prediction errors of all surrogate models on the remaining six systems (AlFe, AlNi, AlTi, CuFe, FeV, NbNi) are consistent, and it is not obvious as to why these systems are harder to learn. 
When generating the data, the same methodology and parameters were used for all alloys, and similar fitting procedures were employed for each surrogate model. 

We point out that whenever the elements that constitute a binary alloy system belong to the same column of the periodic table or are close to each other in the periodic table in terms of atomic number, the surrogate models' predictions are good and vice versa.
Indeed, together these numbers explain 80\,\% of the variance in prediction errors (supplementary material).
A complementary observation is that while \textit{absolute errors} vary from alloy to alloy, \textit{relative errors}($\delta_{\textrm{RMSE}}$), expressed as a percentage of the range of energies of an alloys' subset of the data, remains less than 2.5\% for all systems (Figure \ref{fig:allModels}b).

We trained four of the five investigated surrogate models simultaneously on all 10 alloy systems and compared the mean absolute error (MAE) of these combined models with the average MAE when trained on each alloy system separately (Table~\ref{tab:Train10}; note that RMSE would differ from MAE due to its non-linear nature).  The quantitative agreement indicates that the deviation of the prediction errors is less than 1\,meV/atom when trained on multiple systems.  For the cluster expansion, these results suggest that there is a single set of parameters for generating a prior probability distribution over ECI values (provided in the supporting information) that  works well across a variety of chemistries and lattice types.

For CE, the representation is naturally tied to a particular lattice (e.g. fcc, bcc), making it difficult to train on multiple alloy systems with different lattices at the same time. Here we train a cluster expansion on all alloys by constraining all 30 systems to use a single set of hyperparameters for regularization (i.e. all use the same prior probability distribution of ECI values). The machine-learning surrogate models based on MBTR, SOAP, and MTP do not suffer from the problem of representation being tied to a particular lattice.  They express energy as a continuous function of distances and can be trained on multiple materials simultaneously.

\begin{figure}
	\includegraphics[width=\linewidth]{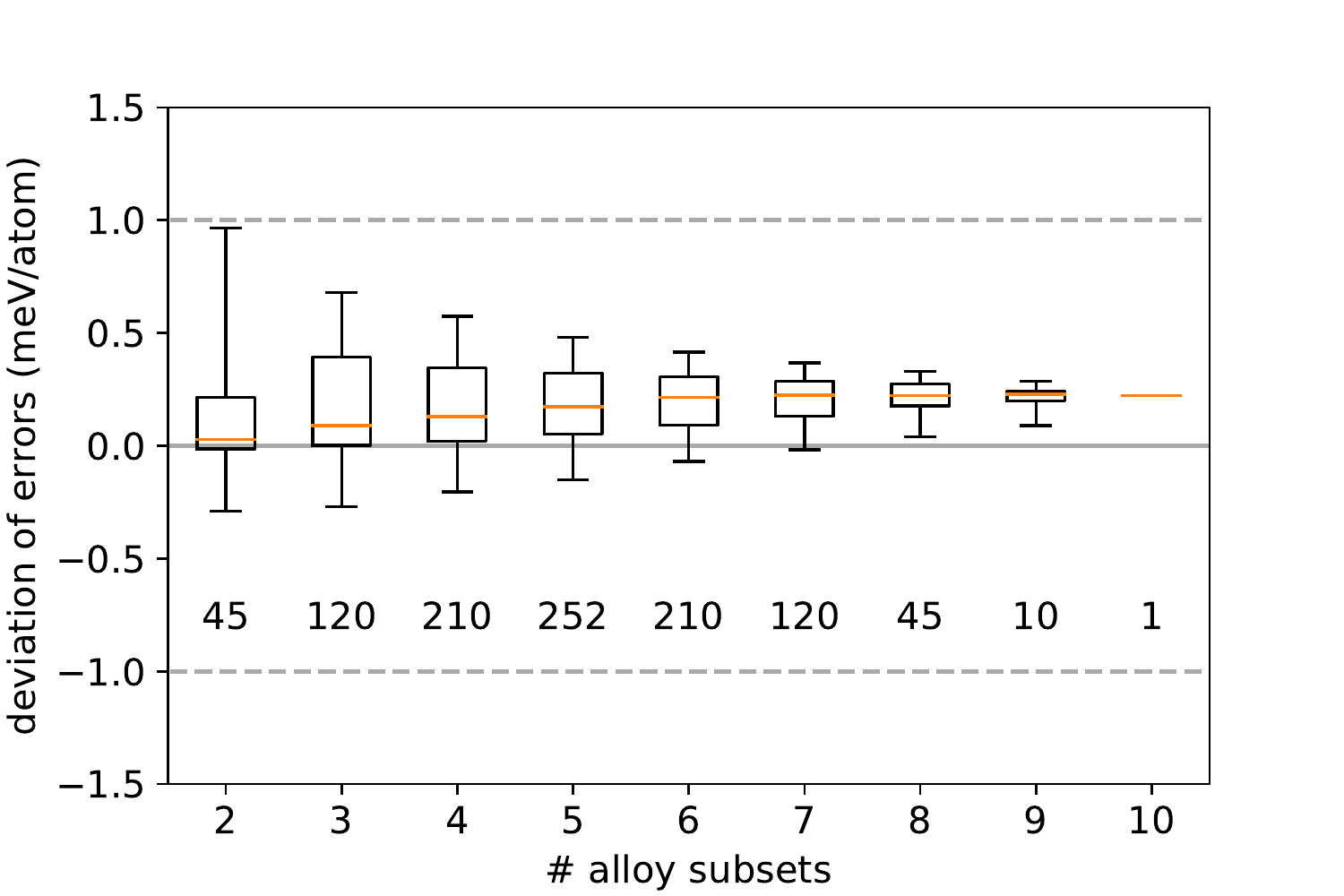}
	\caption{\emph{Performance of MBTR+KRR model for multiple alloy systems.}
		Shown are deviation of mean absolute error (MAE, vertical axis) of an MBTR+KRR surrogate model trained on $k$ (horizontal axis) alloy systems simultaneously from the average MAE of $k$ models trained on each alloy subsystem separately.
		Whiskers, boxes, horizontal line and numbers inside the plot show the range of values, quartiles, median and sample size, respectively.
		Difference in error between individual and combined models is always less than 1\,meV/atom.
	    \label{fig:Deviations}}
\end{figure}

We investigate simultaneous training of alloys in more detail for the MBTR+KRR model.
Figure~\ref{fig:Deviations} presents deviations of the MAE of a single model trained on $k$ alloy systems from the average MAE when the model is trained on each alloy system separately.
In all of the possible $\sum_{k=1}^{10} \binom{10}{k} = 1023$ cases the deviation is below 1\,meV/atom. These deviations are on the order one would expect from minor differences in hyperparameter values. We conclude that prediction errors remain consistently unaffected when increasing the number of simultaneously modeled alloys. 

In the case of MBTR+DNN model, we  observe improvement in prediction errors on the combined model when compared to the average of separate models (Table~\ref{tab:Train10}, Fig. 2 in supplementary material). This suggests that it might be possible to learn element  similarities between chemical element species using a DNN to improve learning rates further.\cite{faber2018alchemical}

\begin{table}[hbtp]
	\caption{\emph{Performance of general models.} 
		Shown are mean absolute errors (MAE) of models trained on all 10 alloy systems simultaneously (right column) versus the average MAE of models trained on individual alloy systems. The combined fit using SOAP+GP was not performed in this work. 
		\label{tab:Train10}}
		
	\medskip
		
	\begin{tabular}{ccc} \hline
		          & \multicolumn{2}{c}{Mean Absolute Errors} (meV/atom)\\ \cline{2-3}
		Surrogate & Average of Sep- & Combined \\
		Model     & arate Models    & Model  \\
		\hline 
		CE       &  4.7  &  4.8 \\ 
		MBTR+KRR &  5.1  &  5.3 \\ 
		MBTR+DNN &  5.1  &  4.6 \\ 
		SOAP+GP  &  4.5  &  ---\\ 
		MTP      &  3.1  &  3.4 \\ 
	\hline

	\end{tabular}

\end{table}

\subsection*{Caveat emptor}

Are reported errors reliable estimates of future performance in applications?
It depends.
We discuss the role of training and validation set composition as an example of the intricacies of statistical validation of machine learning models.

In the limit of infinite independent and identically distributed data, one would simply sample a large enough validation set and measure prediction errors, with the law of large numbers ensuring the reliability of the estimates.
Here, however, data are limited due to the costs of generating them via ab initio simulations, and are neither independent nor identically distributed.
In such a setting, part of the available data is used for validation, either in the form of a hold-out set (as in this work) or via cross-validation, suited for even smaller datasets.

Prediction errors in machine learning models improve with data (otherwise it would not be machine \textit{learning}).
This implies that if only few training samples exist for a ``subclass'' of structures, prediction errors for similar structures will be high.
For example, consider the number of atoms per unit cell in the \texttt{DFT-10B} dataset used here:
There are only 11 structures for each alloy that have 1 or 2 atoms in the unit cell.
Consequently, prediction errors are high for those structures (supplementary material)

In addition to being sparse, smaller unit cells also have a different information content than the larger unit cells. Small unit cells are typically far away from the large unit cells and from each other. Each structure is a point in the representation space and interpolating between structures that are far apart is more prone to error than in regions where the data is tightly clustered. Ideally, the data that the model is trained on would be uniformly distributed in the representation space.  Because small unit cells are few in number and because they have a different information content, it is best to include them in the training set.

For combinatorial reasons, the number of possible structures increases strongly with the number of atoms in the unit cell (Table~\ref{tab:Sizes}).
This biases error statistics in two ways:
As discussed, prediction errors will be lower for classes with more samples.
At the same time, because these classes have more samples, they will contribute more to the measured errors, dominating averages such as the RMSE.

\begin{table}[hbtp]
	\caption{\textit{Size distribution in the \texttt{DFT-10B} dataset.}
		Shown are the number of structures with $k$ atoms in the unit cell, $k \leq 10$ (per alloy; multiply by 10 for the total dataset).
		\label{tab:Sizes}}
		
	\medskip
	
	\begin{ruledtabular}
	\begin{tabular}{lcccccccc} 
		atoms/unit cell &  1  &  2  &   3  &   4  &   5  &    6  &    7  &       8 \\ 
		\#structures    &  4  &  7  &  12  &  48  &  56  &  210  &  208  &  1\,050 \\
	\end{tabular}
	\end{ruledtabular}
\end{table}

\begin{figure}
	\includegraphics[width=\linewidth]{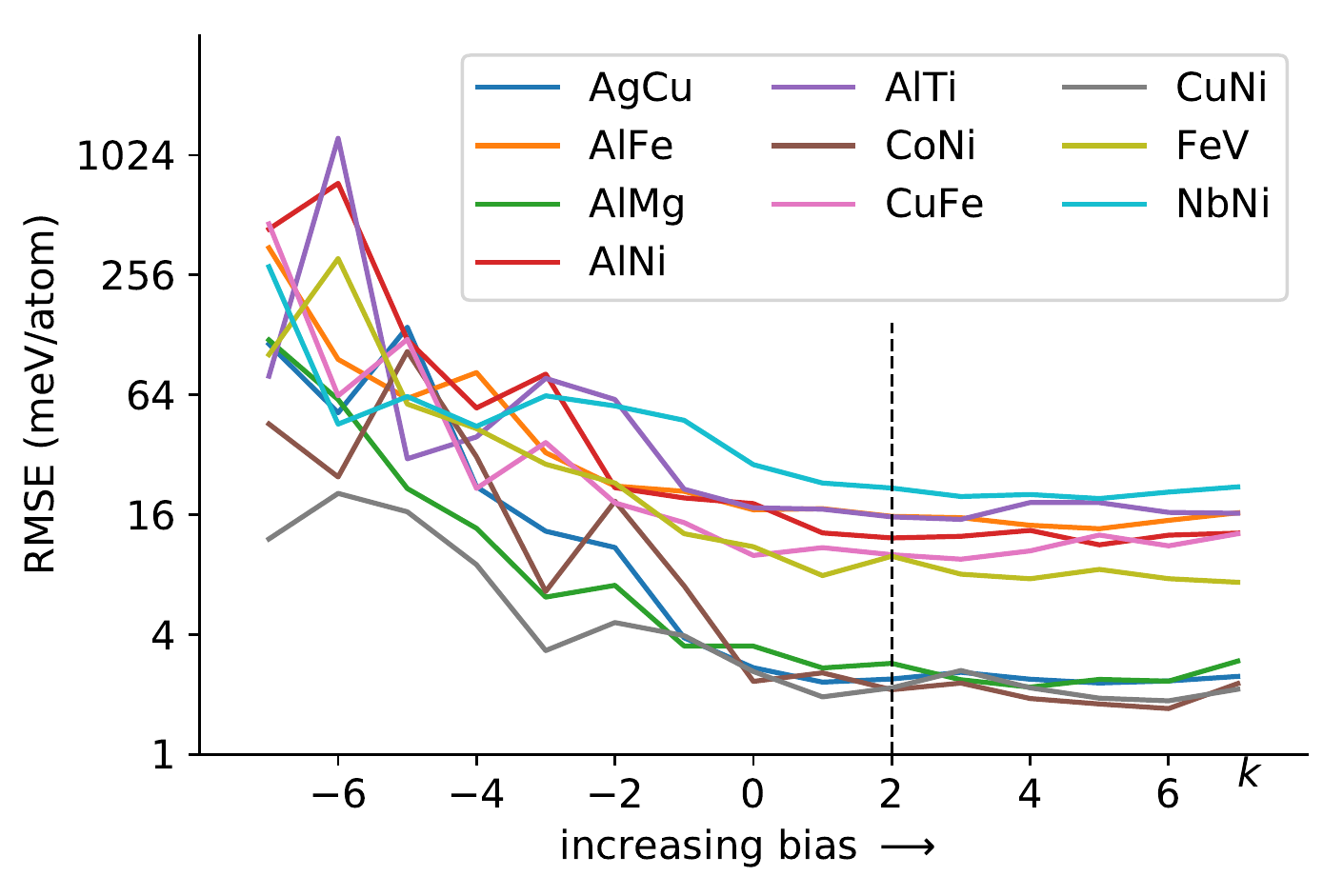}
	
	\caption{\textit{Influence of biased training and validation sets.}
		Shown are the root mean squared errors (meV/atom) as a function of training and validation set composition obtained using MBTR+KRR model.
		See main text for discussion.
		\label{fig:Splits}}
\end{figure}

Figure~\ref{fig:Splits} presents MBTR+KRR prediction errors (RMSE in meV/atom) for different but same-size splits of the data into training and validation sets.
On the left, all structures with $|k|$ or fewer atoms in the unit cell are excluded from the training set (and therefore included in the validation set).
This results in many high-error structures in the validation set, with the effect decreasing for smaller $|k|$.
For $k=0$, size does not influence the split.
On the right, structures with $\leq k$ atoms are always included in the training set, resulting in fewer high-error structures in the validation set.
The dashed line marks the value of $k=2$ recommended in this work (see also supplementary material)

Retrospective errors reported in the literature should, therefore, be critically assessed.
The design of such studies should report on ``representative'' validation sets instead of those tweaked to yield lowest possible errors.
For combinatorial datasets, the smallest structures (those that can be considered to be outliers) should be included in the training set.\cite{rtml2012}

We showed that it is possible to use machine learning to build a combined surrogate model that can simultaneously predict the enthalpy of formation of crystal structures across 10 different binary alloy systems, for three lattice types (fcc, bcc, hcp) and for structures not in their ground state.
In this, we find that the concept of using machine learning to predict  formation enthalpy of materials to be independent of the details of the used surrogate models as predictions of several state-of-the-art materials representations and learning algorithms were found to be in qualitative agreement.
This observation also seems to be congruent with recent efforts towards a unifying mathematical framework for some of the used representations.\cite{wmc2018b}

The ability to use a single surrogate model for multiple systems simultaneously has the potential to simplify the use of surrogate models for exploration of materials spaces by avoiding the need to identify ``homogeneous'' subspaces and then building separate models for each of them.
This also avoids problems such as discontinuities at the boundaries of separate models.

Is it possible to do better?
Recent results suggest that it might be possible to exploit similarities between chemical element species to improve learning rates further.\cite{faber2018alchemical} 
This requires either to explicitly account for element similarities in the representations or to learn element similarities from the data, for example with a DNN.
While such alchemical learning is outside of the scope of this work, we do observe an improvement in prediction errors for the general MBTR+DNN model (Table~\ref{tab:Train10}, Fig. 2 in supplementary material)


\section*{Methods}

\subsection*{Data}

We created a dataset (\texttt{DFT-10B}) containing structures of the 10 binary alloys AgCu, AlFe, AlMg, AlNi, AlTi, CoNi, CuFe, CuNi, FeV, NbNi. 
Each alloy system includes all possible unit cells with 1--8 atoms for face-centered cubic (fcc) and body-centered cubic (bcc) crystal types, and all possible unit cells with 2--8 atoms for the hexagonal close-packed (hcp) crystal type. This results in 631 fcc, 631 bcc and 333 hcp structures, yielding 1595 $\cdot$ 10 = 15\,950 unrelaxed structures in total. We refer to this dataset as \texttt{DFT-10B} in this work. The cell shape, volume, and atomic positions were not optimized and the calculations are all unrelaxed, for the sake of efficiency. The crystal structures were generated using the enumeration algorithm by Hart and Forcade \cite{hf2008}. 

Lattice parameters for each crystal structure were set according to Vegard's law. \cite{v1921,da1991}
Total energies were computed using density functional theory (DFT) with projector-augmented wave (PAW) potentials \cite{kj1999,b1994,kh1994} within the generalized gradient approximation (GGA) of Perdew, Burke, and Ernzerhof \cite{pbe1996} (PBE) as implemented in the Vienna Ab Initio Simulation Package \cite{kf1996,kf1996b} (\texttt{VASP}).
The $k$-point meshes for sampling the Brillouin zone were constructed using generalized regular grids.\cite{wmm2016,morgan2018efficiency} The details of the $k$-point density for all 10 alloys is mentioned in the supplementary material (table I)

\subsection*{Models}

All single-alloy surrogate models were trained using the same set of 1000 randomly selected crystal structures,  including optimization of hyperparameters, and the prediction errors are reported on a hold-out test set of 595 different structures, never seen during training. The same training and test structures are used for all binaries. Models trained on multiple alloys use the union of the individual alloy's splits.
Parametrization details of all surrogate models used in this work can be found in the supplementary material.

\section*{Acknowledgments}

CN is thankful to Kennedy Lincoln and Wiley Morgan for insightful discussions.  
CN, BB, CR, and GH acknowledge the funding from ONR  (MURI N00014-13-1-0635). MR acknowledges funding from the EU Horizon 2020 program Grant 676580, The Novel Materials Discovery (NOMAD) Laboratory, a European Center of Excellence.
AS was supported by the Russian Science Foundation (Grant No 18-13-00479). TM acknowledges funding from the National Science Foundation under award number DMR-1352373 and computational resources provided by the Maryland Advanced Research Computing Center (MARCC).

\section*{Data availability}

The dataset (\texttt{DFT-10B}) generated and used for the current work is publicly available at \url{https://qmml.org}.

\section*{AUTHOR CONTRIBUTIONS}
C.N. conceived the idea, generated the dataset, ran the calculations of the MBTR-based models, interpreted the results, and wrote a significant portion of the paper. M.R. was responsible for dataset analysis, did the MBTR+KRR calculations, and also wrote a significant portion of the paper. B.B. helped generate the dataset and analyzed the MBTR+KRR calculations. A.V.S. performed the MTP calculations. T.M. performed all cluster expansion calculations. C.W.R. performed SOAP+GPR calculations. G.C. provided guidance and expertise in applying SOAP to our dataset. D.W.W provided his expertise for the MBTR+DNN model. G.L.W.H. contributed many ideas and critique to help guide the project and helped write the paper.

\section*{Competing interests}
The authors declare that there are no competing interests.


\section*{Supplementary material}

\subsection{Method details}

\subsubsection{Cluster expansion}

To determine the cutoff distances for the cluster expansions and determine the initial parameters for the prior probability distributions, we used a length scale in which the edge of a bcc unit cell is 1 unit of length and assumed the hcp, bcc, and fcc crystal structures all had the same nearest-neighbor distance.  The cutoff distances used to determine the set of clusters included in the expansion are as follows:

\begin{table}[H]
\centering
\begin{tabular}{ll}
\hline
Number of sites in cluster & 
Maximum distance between sites \\ \hline
2 & 8 \\
3 & 4\\
4&2\\
5&1.5\\
6&1.5\\
\hline
\end{tabular}
\end{table}

This resulted in a total of 791, 941, and 2870 distinct orbits of clusters in the bcc, fcc, and hcp expansions, respectively, including the empty cluster.  These numbers were reduced after fitting by ``trimming" the cluster expansions, in which cluster functions with very small ECI were removed from the expansion.  To determine which clusters to remove, we used the fact that when the cluster functions are orthonormal, the expected squared error due to truncation, $E(\textrm{error}^2)$ , is given by

\begin{equation}
E(\textrm{error}^2)=\sum_b V_b^2,
\end{equation}

\noindent where $V_b$  is the ECI for the  $b$-th cluster function, the expectation of the squared error on the left is over all possible lattice decorations, and the sum on the right is over cluster functions excluded from the expansion.  Thus removing an orbit of clusters with multiplicity  $m_b$ increases the expected squared error by $m_bV_b^2$ .  To trim clusters from the expansion with little loss of accuracy, we removed all orbits of cluster functions for which $\sqrt{ m_bV_b^2} < 10^{-5}$ eV.  The trimming procedure changed the final average root-mean-squared prediction errors on the training sets by less than  $10^{-5}$ eV / atom and removed on average more than 70\% of the ECIs in the expansions. 

The ECIs for the cluster expansions were fit to the training data using the Bayesian approach with a multivariate Gaussian prior distribution \cite{mc2009b}. The inverse of the covariance matrix for the prior, $\Lambda$, was diagonal, with elements given by

\begin{equation}
\lambda_{\alpha \alpha} =
\left\{
	\begin{array}{ll}
		 0  & \mbox{for }   n_{\alpha} = 0 \\
		e^{-\lambda_1} & \mbox{for } n_{\alpha} = 1 \\
		e^{-\lambda_2}e^{-\lambda_3 r_\alpha}n_{\alpha}^{\lambda_4} & \mbox{for } n_\alpha > 1
	\end{array}
\right\},
\end{equation}

\noindent where $n_\alpha$ is the number of sites in cluster function $\alpha$ and $r_\alpha$  is the maximum distance between sites in Angstroms.  The parameters $\lambda_1$, $\lambda_2$, $\lambda_3$, and $\lambda_4$  were initially set to 10, 10, 5, and 5 respectively then optimized by using a conjugate gradient algorithm to minimize the root mean square leave-one-out cross-validation error, an estimate of prediction error\cite{wc2002}. For the combined fit, in which a single set of regularization parameters were used for all 30 cluster expansions, the optimized values of $\lambda_1, \lambda_2, \lambda_3,$ and $\lambda_4$ were 10.0, 
20.8, 4.2, and 15.3 respectively.

\subsubsection{MBTR+KRR}

The Many-Body Tensor Representation (MBTR) numerically represents atomistic systems as distributions of many-body terms, such as atom counts, distances, and angles, stratified (separated) by chemical elements.
For details please consult Ref.~\cite{hr2017}.
Kernel ridge regression \cite{r2015} with a Gaussian kernel was employed throughout.
In this work, we use the following parametrization:

\medskip

\noindent
\begin{table}
\centering
\begin{tabular}{@{}cllll@{}} \hline
	$k$ & geometry & weighting & discretization & $\sigma$ \\ \hline
	2   & \texttt{1/distance} & \texttt{identity{\textasciicircum}2} & $(0, 0.005, 90)$        & $2^{-17}$  \rule{0pt}{2.5ex} \\
	2   & \texttt{1/distance} & \texttt{identity{\textasciicircum}2} & $(0, 0.005, 90)$        & $2^{-4.5}$ \\
	3   & \texttt{angle}      & \texttt{1/dotdotdot}                 & $(-0.15, \pi/100, 100)$ & $2^{-14}$  \\ \hline
\end{tabular}
\end{table}
\medskip

We did not use 1-body terms as enthalpies of formation are the result of a linear operation in atom counts already.
Values for the $\sigma$ hyperparameter above refer to Fig. 3 in manuscript, where we used fixed hyperparameter values (Gaussian kernel $\sigma = 2^7$, KRR regularization strength $\lambda = 2^{-20}$).
For individual models, hyperparameters were optimized on a base-2 logarithmic grid.

\subsubsection{MBTR+DNN}

The mathematical details of the many-body tensor representation for the crystal structures are mentioned in ref. \cite{hr2017}.  Each crystal structure is expanded in terms of distributions ($k$-body terms) of atom counts, (inverse) distances and  angles. The Gaussian kernel with  a variance ($\sigma$) of $11.3$ was used for fitting. Each MBTR vector is 1450 long and was optimized using a grid search. The details of the weighting functions, smearing parameters for each $k$-body term are as follows,

\medskip

\noindent
\begin{table}
\centering
\begin{tabular}{@{}cllll@{}} \hline
	$k$ & geometry & weighting & discretization & $\sigma$ \\ \hline
	1   & \texttt{atom count} & \texttt{1/identity} & $(0.5, 1, 25)$        &  $10^{-4}$ \rule{0pt}{2.5ex} \\
	2   & \texttt{1/distance} & \texttt{identity{\textasciicircum}2} & $(0.1 , 0.005, 70)$        & $ 2^{-17}$ \\
	3   & \texttt{angle}      & \texttt{1/dotdotdot} & $(0.1, 0.05, 140)$ & $2^{-8.}$ \\ \hline
\end{tabular}
\end{table}
\medskip

MBTR+DNN model uses the same parameters as MBTR+KRR model for generating the representation. The only difference  between the representations is that the $k$-body terms in MBTR+DNN model are stratified by all 10 elements instead of just two.  This results in a representation vector which is 147100 long. The architecture of the convolution neural network used in this work is listed in the table below. 

\begin{table}[H]
\centering
\begin{tabular}{ll}
\hline \\
Layer type &  Specifications \\
\hline \\
Fully connected layer & (Size: 2048) \\
Fully connected layer & (Size: 1024) \\
 Reshaping data  & (Size: 4 x 4 x 64) \\
Convolution transposed layer & (Kernel: 5 x 5, 64 filters) \\
Convolution layer & (Kernel: 3 x 3, 64 filters) \\
Max pooling layer & (Pool size: 2 x 2; stride: 2 x 2) \\
Convolution layer & (Kernel: 3 x 3, 32 filters) \\
 Reshaping data  & (Size: 1 x 1024) \\
Fully connected layer & (Size: 128) \\
Fully connected layer & (Size: 64) \\
Fully connected layer & (Size: 4) \\
Fully connected layer & (Output; size: 1) \\
\hline
\end{tabular}
\end{table}

The DNN code is implemented using the Tensorflow framework (software available from \texttt{www.tensorflow.org}).  The models were trained with a mini-batch size of 50 and the RMSE error is used as the cost function for optimizing the weights of the network.

\subsubsection{SOAP+GP}

GAP fits were generated for each alloy system using a 2-body + SOAP
approach. The standard deviation (SD, parameter $\delta$) of the
Gaussian process for the 2-body GAP is set to match the SD in energies
of the training set. After fitting the 2-body potential, another SOAP
GAP is fit with its SD set to match the remaining RMSE of the 2-body
GAP relative to the DFT energies in the training set. The fits were
performed using \verb|teach_sparse|  (software available from \texttt{www.libatoms.org}) with the following
parameters for the 2-body GAP:


\begin{table}[H]
\centering
\begin{tabular}{lc}
\hline \\
Parameter &  Value \\
\hline \\
Cutoff & 6.0 \AA \\
Sparse points & 10 \\
\hline
\end{tabular}
\end{table} 

\noindent and for SOAP, parameters were set to:

\begin{table}[H]
\centering
\begin{tabular}{lc}
\hline \\
Parameter &  Value \\
\hline \\
Cutoff & 4.5 \AA \\
Sparse points & 500 \\
$l_{\mathrm{max}}$ & 8 \\
$n_{\mathrm{max}}$ & 8 \\
$\zeta$ & 2 \\
$\sigma_{\mathrm{atom}}$ & 5 \\
\hline
\end{tabular}
\end{table} 

As described above, $\delta$ is set using the standard deviation of
the Gaussian Process based on the training set for the 2-body and SOAP
fits respectively. The following table lists these values for each of
the alloy systems.

\begin{table}[htb]
\centering
\begin{tabular}{lcc}
\hline \\
Parameter & $\delta$ (2-body) & $\delta$ (SOAP) \\
\hline \\
AgCu & 0.43 & 0.0126 \\
AlFe & 0.84 & 0.044 \\
AlMg & 0.43 & 0.0126 \\
AlNi & 0.396 & 0.0193 \\ 
AlTi & 0.78 & 0.03 \\
CoNi & 0.27 & 0.0337 \\
CuFe & 0.84 & 0.0446 \\
CuNi & 0.315 & 0.0207 \\
FeV & 0.184 & 0.0407 \\
NbNi & 0.9 & 0.05\\
\hline
\end{tabular}
\end{table} 

For all alloy fits, the error hyperparameter $\sigma$ was set to 1
meV for energies. Force and virial were not used in the fits. Because
pure energies have a large effect on the predicted formation
enthalpies, we increased the error hyperparameter to $10^{-4}$ for
those training configurations that represented pure elements. This
ensured accurate reproduction of the pure energies so that enthalpy
errors closely match errors in total energy for configurations.

The parameter $\epsilon_0$ was calculated for each isolated atom by
including a padding of 10\, \AA\, around a single atom and using the same
pseudopotential as the bulk calculations discussed above. These
energies were converged with respect to basis set size and used only
the $\gamma$ k-point.

\subsubsection{MTP}

MTP was introduced in Ref.\ \cite{s2016} for single-component system and in Refs.\ \cite{gubaev2018machine,gubaev2018-mtp-multicomponent} was extended to multicomponent systems.
MTP partitions the predicted energy into contributions of environments of each atom.
Around the central atom of an environment, the neighboring atoms form shells.
In these shells, atoms are assigned fictitious weights depending on the distance to the central atom, their types, and the type of the central atom.
These weights are free parameters fitted from data.
An environment is described by moments of inertia of these shells.
All possible contractions of one or more moment tensor to a scalar comprise an infinite sequence of basis functions.
This sequence is truncated to yield a finite set of basis functions used in a particular MTP model.
The contribution of an environment to the energy is, thus, a linear combination of basis function with coefficients which are also found from data.
Refer to Ref. \cite{gubaev2018-mtp-multicomponent} for more details.

In this work for binary systems, we used an MTP with about 300 basis functions.
The cutoff for atomic environments was 7 \AA.
The environments were described by five shells, and the dependence of the weight of a neighbor on the distance to the central atom of the environment was described by eight basis functions.
Thus, the total number of parameters in a binary MTP is $5\times 8\times 2^2+300 \approx 450$ (the factor $2^2$ follows from the fact that there can be two types of the central atom and two types of each neighboring atoms).
For the 10-component MTP, we used six shells and 850 basis functions, totaling  about $6\times 8\times 30+864 \approx 2300$ parameters, where the factor 30 is the number of interacting pairs of atoms.

\subsection{Dataset details}

\begin{table}[h]
\caption{\emph{\textit{k}-point density} Shown are the minimum and maximum values of \textit{k}-point density across all structures for each of the alloys for computing the DFT total energies.}
\label{kDensity}
\begin{tabular}{ccc} \hline \\ [-1.6ex]
		          & \multicolumn{2}{c}{Number of \textit{k}-points}/$\mathrm{\AA}^{3}$\\ \cline{2-3} \\  [-1ex]
		System     & Maximum   & Minimum  \\
		 \hline \\  [-1ex]
AgCu & 550 & 516 \\
AlFe & 596 & 468 \\
AlMg & 635 & 478\\
AlNi & 589 & 464 \\ 
AlTi & 535 & 399 \\
CoNi & 554 & 433 \\
CuFe & 568 & 444\\
CuNi & 561 & 440 \\
FeV &  480 & 401 \\
NbNi &516 & 472\\	
\hline
\end{tabular}
\end{table}

\subsection{Analysis of dataset and models}

\begin{figure}[h]
	\includegraphics[scale=0.6]{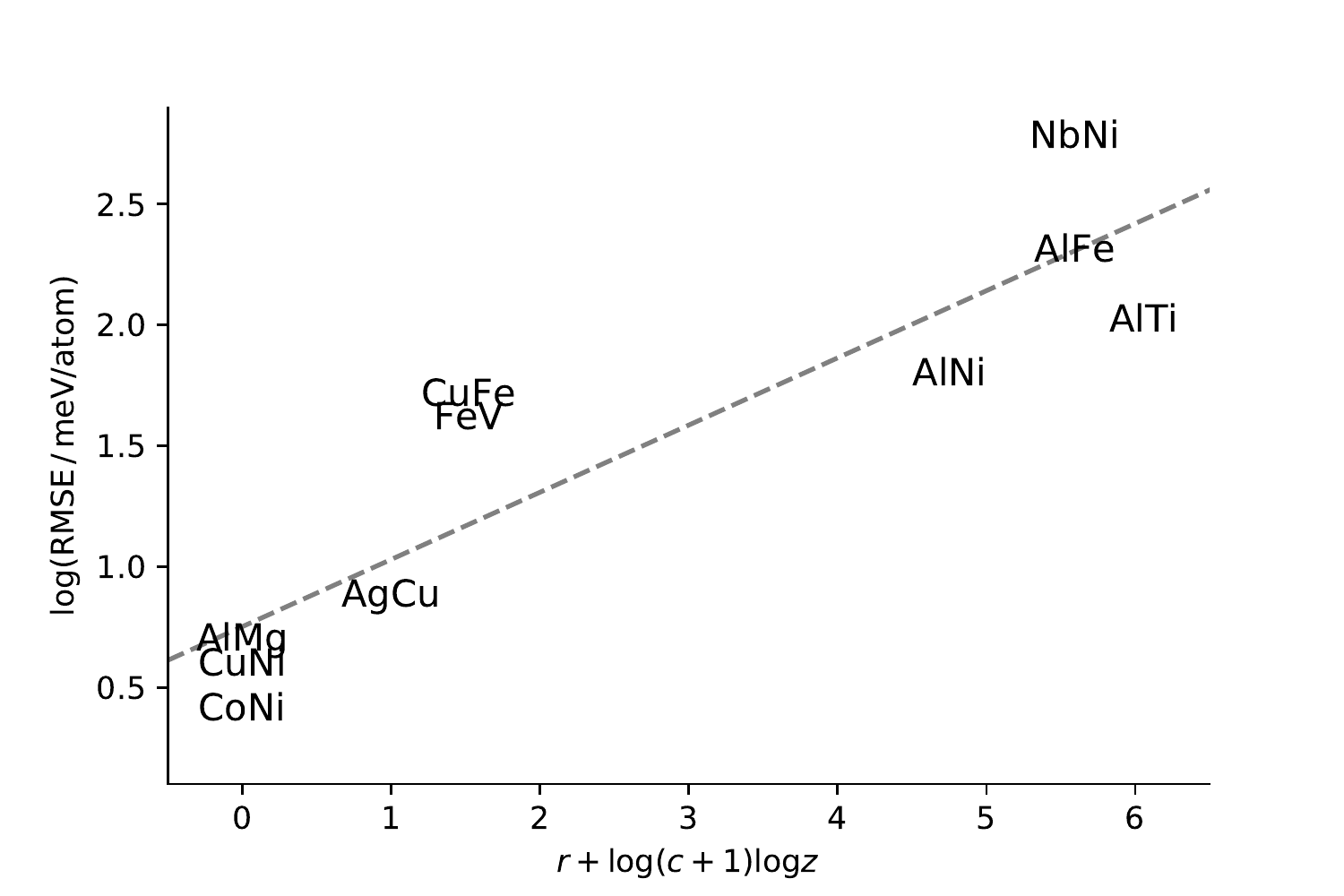}
	\caption{\emph{Alchemical similarity explains prediction errors.}Shown are the logarithmized root mean squared error (RMSE; compare Fig. 2 as a function of an analytic expression in the difference in row $r$ and column $c$ of the periodic table as well as atomic number $z$ of the two chemical element species of a binary alloy.
		$R^2 = 0.81$.
		\label{fig:AlchCorr}}
\end{figure}

\begin{figure*}[!htb]
	\includegraphics[width=\linewidth]{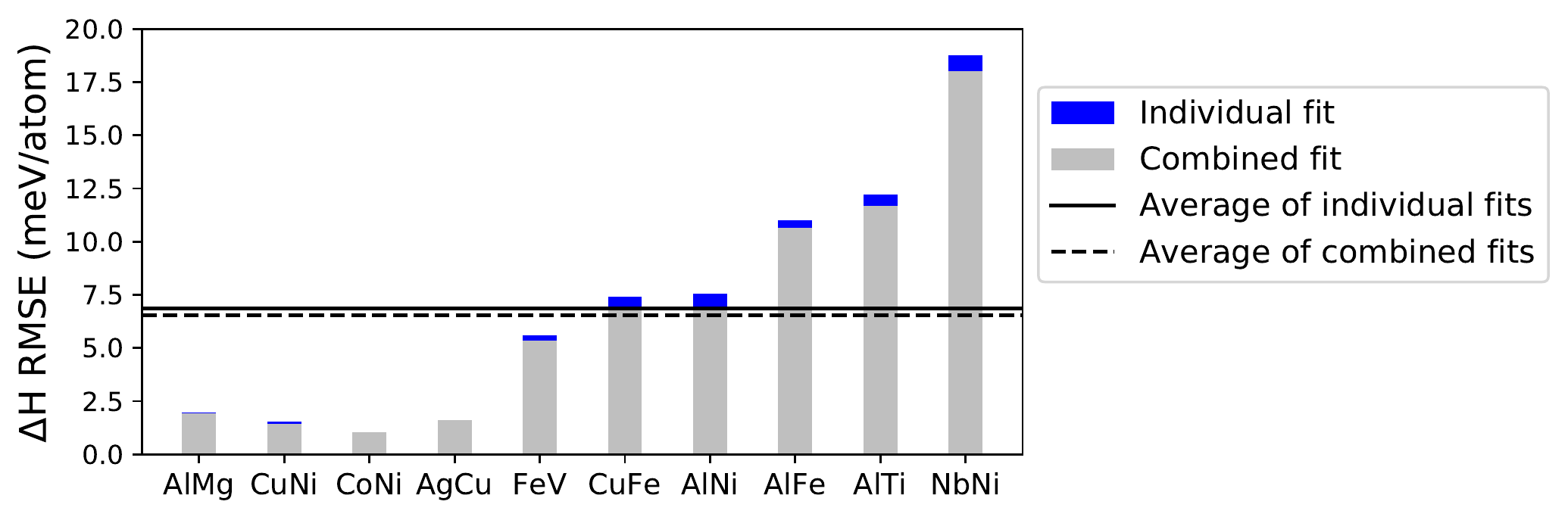}
	\caption{\emph{Improvement of MBTR+DNN model on all alloys.}
		Shown are the root mean squared error (RMSE) when trained on each alloy separately (blue bars) and on all alloys simultaneously (grey bars).
		\label{fig:CrossTalk}}
\end{figure*}

\begin{figure*}[phtb]
	\includegraphics[width=\linewidth]{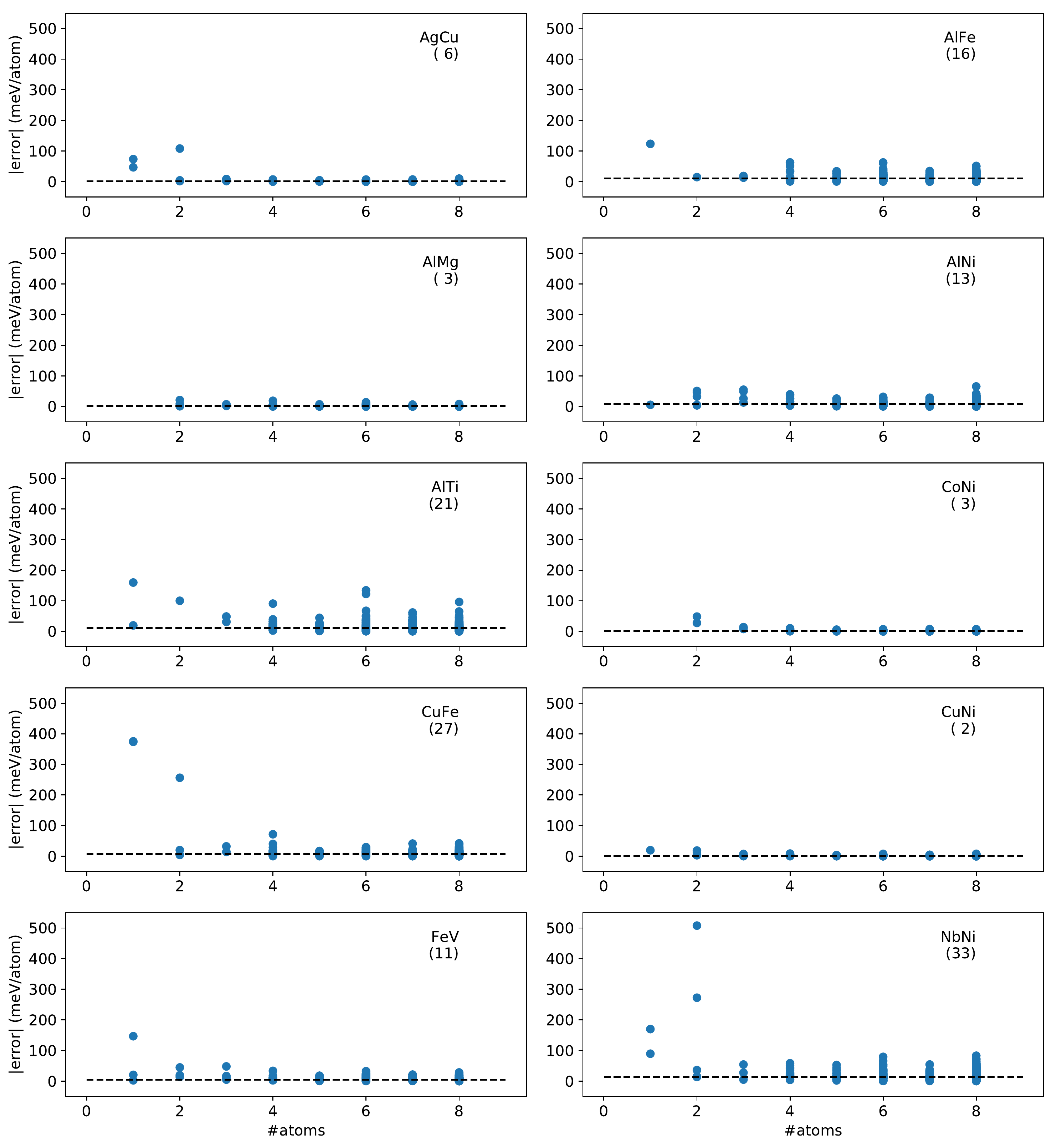}
	\caption{\textit{Influence of unit cell size on errors}.
		Shown are the absolute errors (meV/atom) as a function of the number of atoms in the unit cell for a validation set of 595 randomly chosen structures using the MBTR+KRR model.
		The number in brackets and the dashed line indicate the root mean squared error (RMSE, meV/atom) and the median absolute error (meV/atom) on the same set.
		If small structures (one or two atoms in the unit cell) are not contained in the training data (that is, are shown in the plot) they tend to have larger errors, increasing overall RMSE as well.
		If all small structures are contained in the training data, the overall RMSE is low (AlMg, CoNi).
		Retraining models with small structures included in the training set improved RMSE in all cases, by an amount depending on how many structures were added.
		\label{fig:SizeError}}
\end{figure*}

\begin{figure*}[phtb]
	\includegraphics[width=\linewidth]{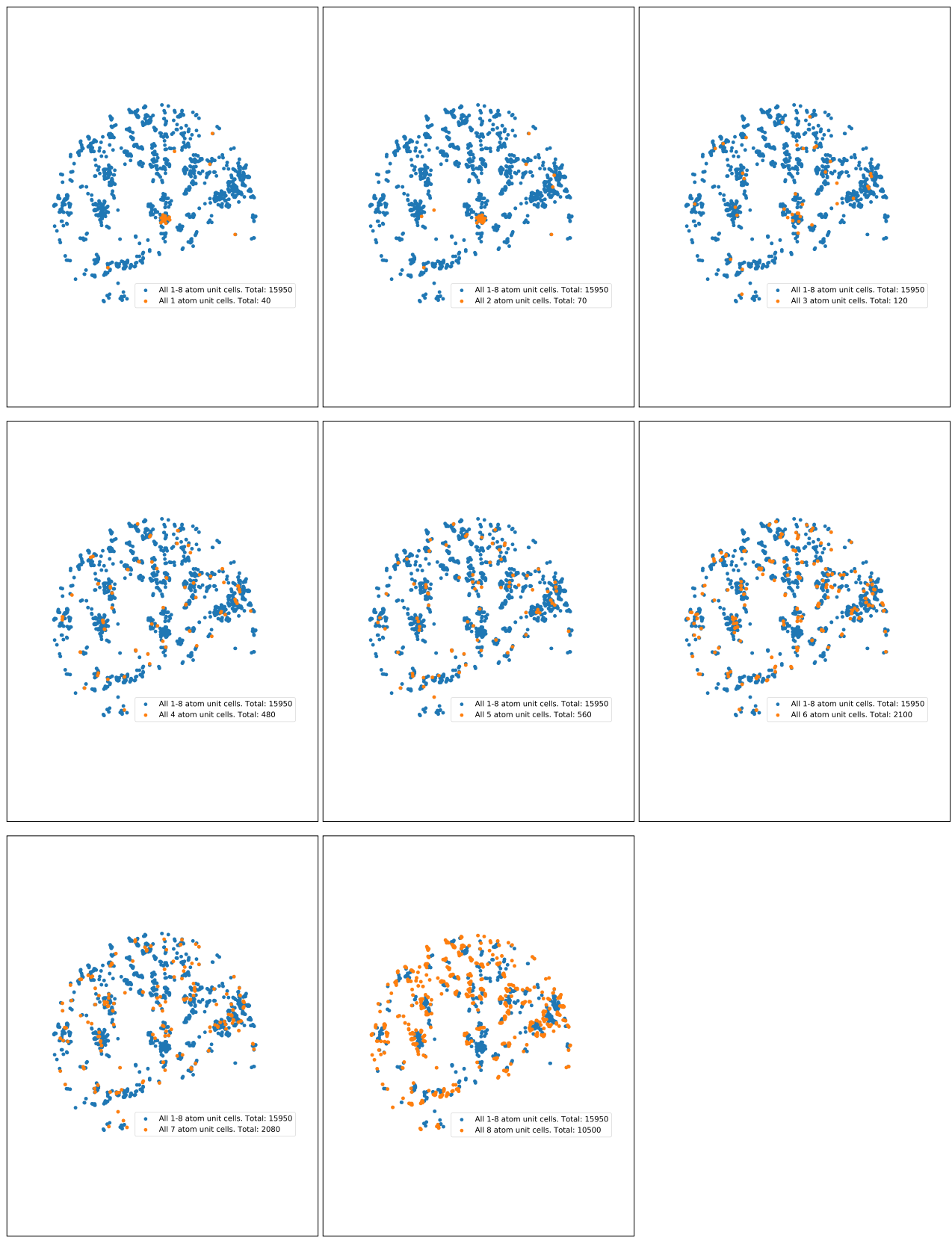}
	\caption{Visualizing all 15950 structures (\texttt{DFT-10B}) using a t-SNE plot. Each structure in the higher-dimensional space (MBTR) is graphically represented on a 2D plane using t-SNE\cite{maaten2008visualizing} method. We can observe that 1 or 2 atom unit cells are not representative of larger unit cells in the dataset and are away from other higher atom unit cells. This is a possible reason for high prediction errors when 1 or 2 atom cells are not included in the training set.}
		\label{fig:until_k_at}
\end{figure*}

\clearpage
\section*{References}


\bibliographystyle{unsrt}
\bibliography{SMFM}

\end{document}